\documentclass[11pt,twoside]{article}
\usepackage{asp2014}
\usepackage[dvipsnames]{xcolor}
\usepackage{ulem}

\aspSuppressVolSlug
\resetcounters

\begin{document}

\title{Chemo-kinematic analysis of metal-poor stars with unsupervised machine learning}

% full name: Andr\'{e} Rodrigo da Silva
\author{Andr\'{e}~R.~da~Silva$^1$, Rodolfo~Smiljanic$^1$, and Riano~E.~Giribaldi$^1$}
\affil{$^1$Nicolaus Copernicus Astronomical Center, Polish Academy of Sciences, ul. Bartycka 18, 00-716, Warsaw, Poland; \email{arodrigo@camk.edu.pl}}
% remove/add as you need

% remove/add authors as you need
\paperauthor{Andr\'{e}~R.~da~Silva}{arodrigo@camk.edu.pl}{0000-0002-7758-656X}{Nicolaus Copernicus Astronomical Center, Polish Academy of Sciences}{}{Warsaw}{}{00-716}{Poland}
\paperauthor{Rodolfo~Smiljanic}{rsmiljanic@camk.edu.pl}{0000-0003-0942-7855}{Nicolaus Copernicus Astronomical Center, Polish Academy of Sciences}{}{Warsaw}{}{00-716}{Poland}
\paperauthor{Riano~E.~Giribaldi}{riano@camk.edu.pl}{0000-0002-9420-560X}{Nicolaus Copernicus Astronomical Center, Polish Academy of Sciences}{}{Warsaw}{}{00-716}{Poland}
% remove/add as you need

% leave these next few aindex lines commented for the editors to enable them. Use Aindex.py to generate them for yourself.
% first presenting author should be the first entry for bold-facing the author index page-reference
%\aindex{da Silva,~A.R.}
%\aindex{Smiljanic,~R.}
%\aindex{Giribaldi,~R.E.}
% remove/add as you need

% leave the ssindex lines commented for the editors to enable them, use Index.py to suggest yours
%\ssindex{catalogues!Gaia Data Release 2}
%\ssindex{catalogues!GALAH Data Release 2}
%\ssindex{algorithm!machine learning!unsupervised}
%\ssindex{methods!statistical!unsupervised clustering}

% leave the ooindex lines commented for the editors to enable them, use ascl.py to suggest yours
%\ooindex{STARS, ascl:1107.008}
%\ooindex{GALAXY, ascl:1904.002}
%\ooindex{GAIA, ascl:1403.024} 
  
\begin{abstract}

Metal-poor stars play an import role in the understanding of Galaxy formation and evolution. Evidence of the early mergers that built up the Galaxy might remain in the distributions of abundances, kinematics, and orbital parameters of the stars. In this work, we report on preliminary results of an on-going  chemo-kinematic analysis of a sample of metal-poor ([Fe/H] $\leq$ -1.0) stars observed by the GALAH spectroscopic survey. We explored the chemical and orbital data with unsupervised machine learning (hierarchical clustering, k-means cluster analysis and correlation matrices). Our final goal is to find an optimal way to separate different Galactic stellar populations and stellar groups originating from merging events, such as Gaia-Enceladus and Sequoia.

%We cross-matched this sample with the \textit{Gaia} DR2 catalog to obtain a final sample of 1\,072 objects with precise parallaxes and proper motions, radial velocities, and with metallicities [Fe/H] $\leq$ $-$1.0 dex. With this selection, we used \textit{galpy} to integrate stellar orbits and derive parameters such as orbital eccentricity, farthest distance reached from the plane, action angles and angular momentum. 
  
\end{abstract}

\section{Introduction}

Galaxies grow through the accretion and merging of smaller stellar systems \citep[e.g.,][]{BlandHawthorn2014}. It is sometimes possible to identify signatures of these early mergers as chemo-kinematic stellar substructures. In many instances, however, the criteria used to separate and study such substructures is arbitrarily defined by carving boxes out of the parameter space. In this work, we are investigating ways to improve the selection criteria using unsupervised machine learning, and trying to account for the possible distributions of stellar properties of the objects belonging to each substructure.

%The field of Galactic Archaeology has developed with the aim of reconstructing the history of the formation and evolution of the Milky Way using large stellar surveys.

The first unequivocal evidence of mergers was the discovery of the Sagittarius dwarf galaxy by \citet{Ibata1994}. Since then, others have discovered more pieces of the Galactic merger puzzle. Recently, \citet{Helmi2018} and \citet{Belokurov2018} independently discovered a possible merger that occurred in the Milky Way about 8-11 Gyr ago. The object that merged with the Galaxy, often called Gaia-Enceladus or Gaia-Sausage (GES), could be the origin of the inner halo and the thick disk of the Milky Way \citep{Helmi2018}. The second data release (DR2) of \textit{Gaia} \citep{Gaia} has helped in the identification of additional merger candidates such as the Sequoia \citep{Myeong2018} and the Thamnos substructures \citep{Koppelman2019}.
%, Aleph, Arjuna, I'itoi and Wukong \citep{Naidu2020}. 

\section{Data and analysis}

In this work, we used the proper motions and parallaxes of \textit{Gaia} DR2. Chemical parameters were obtained from the DR2 of GALAH \citep[Galactic Archaeology with Hermes,][]{Galah}. GALAH delivered chemical abundances of 26 chemical elements for $\sim$340\,000 stars from the analysis of spectra with R $\sim$ 28\,000. We cross-matched metal-poor stars from the GALAH survey (selected with [Fe/H] $\leq$ $-$1.0, 2.0\,$\leq\log g\leq$\,5.0, $T_{\rm eff}$ $\geq$\,4\,kK and with cannon\_flag = 0) with \textit{Gaia} (selecting those with $\sigma_\varpi / \varpi <$ 20\%). This resulted in 1\,072 stars.\par
Using \textit{Gaia} proper motions and GALAH radial velocities, we then integrated the stellar orbits using \textit{galpy} \citep{Bovy2015} for a period of 10 Gyr using the Milky Way potential from \citet{McMillan2017}. From \textit{galpy}, we extracted the action angles (J$_r$, J$_\phi$,J$_z$), two integrals of motion (E, L$_z$), eccentricity, cylindrical radius, and other orbital parameters. We also performed Monte Carlo simulations in order to estimate the uncertainties.

%\section{Unsupervised Machine Learning}
t-distributed stochastic neighbor embedding (t-SNE) is a manifold learning method that uses affinity between the data points as probability. This is particularly useful for exploring structures from an N dimensional problem in a 2D map. The location of the points in this map is given by minimizing the Kullback-Leibler divergence. For the analysis that we report here, we used t-SNE to identify the stars in our sample that are close together in the parameter space. The quantities used were: J$_r$, J$_\phi$, J$_z$, and the orbital eccentricity, together with the chemical parameters [Fe/H] - representing iron-peak elements, [Mg/Fe] - representing the $\alpha$-elements, [Ba/Fe] - representing the s-process elements and [Eu/Fe] - representing the r-process elements.\par
In order to decide which is the best number of cluster components for these data, we did a k-means clustering analysis with increasing number of clusters. The first number of clusters past the mark of 95\% of distortion explanation was considered as the optimal number (in this case nine). We then used agglomerative clustering using Ward hierarchical method to separate the t-SNE map into the nine clusters. We used the python module scikit-learn \citep{pedregosa2011} to perform this analysis. 

\section{Discussion}
In Fig. \ref{fig1}, the t-SNE map is shown with the color-coded clusters. The definition of structures associated to mergers have in general been done using only kinematical parameters  \citep[see][]{Helmi2018,Belokurov2018,Myeong2018,Massari2019}, with a few exceptions like in \citet{Naidu2020}. \citet{Massari2019} and \citet{Myeong2018} disagree on criteria used to classify the GES and Sequoia stars (see Fig. \ref{fig2}). Figure \ref{fig3} helps to compare which proposed Galactic substructures found in the literature is closer to each of the clusters that we found.

Comparing the results of our analysis with those of the literature, we find that: in gray scale, we have the stellar groups that correspond to what is generally agreed to be the disk of the Galaxy; in pink scale is the GES close to the limits defined by \citet{Massari2019}; the red points agree with \citet{Myeong2018} and \citet{Naidu2020} for the Sequoia substructure; the stars in blue are intermediary between the disk and the halo of the Milky Way; and the lighter blue cluster is on the same region as \citet{Naidu2020} defines the Wukong substructure (as a prograde, metal-poor and $\alpha$-rich substructure). In Fig. \ref{fig4}, it is possible to separate the pink shaded group from the gray. The dark blue cluster is on the same region of what \citet{Naidu2020} called a ``metal-weak thick disk''. To summarize, we found that unsupervised machine learning techniques are powerful tools that can help to recover halo substructures associated to early merging events of the Milky Way.

\articlefigure[width=\textwidth, height=.3\textheight]{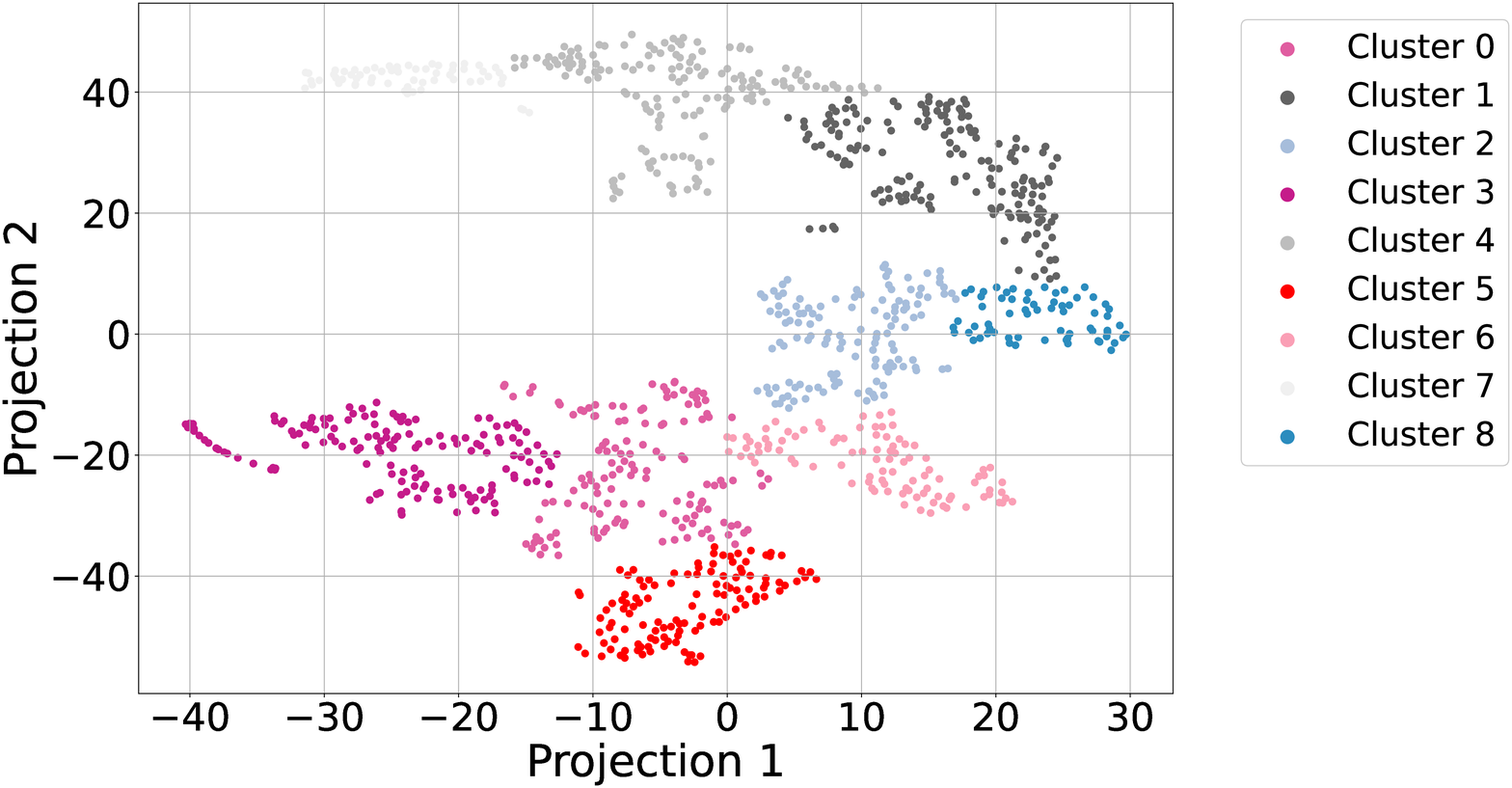}{fig1}{t-SNE projection of the parameter space for J$_\phi$, J$_r$, J$_z$, [Fe/H], [Mg/Fe], [Ba/Fe], [Eu/Fe], orbital eccentricity. Gray-colored symbols represent the disk, in pink is the Gaia-Enceladus-Sausage (GES), in red is Sequoia and in blue the thick disk transition to the halo.}

\articlefigure[width=.9\textwidth]{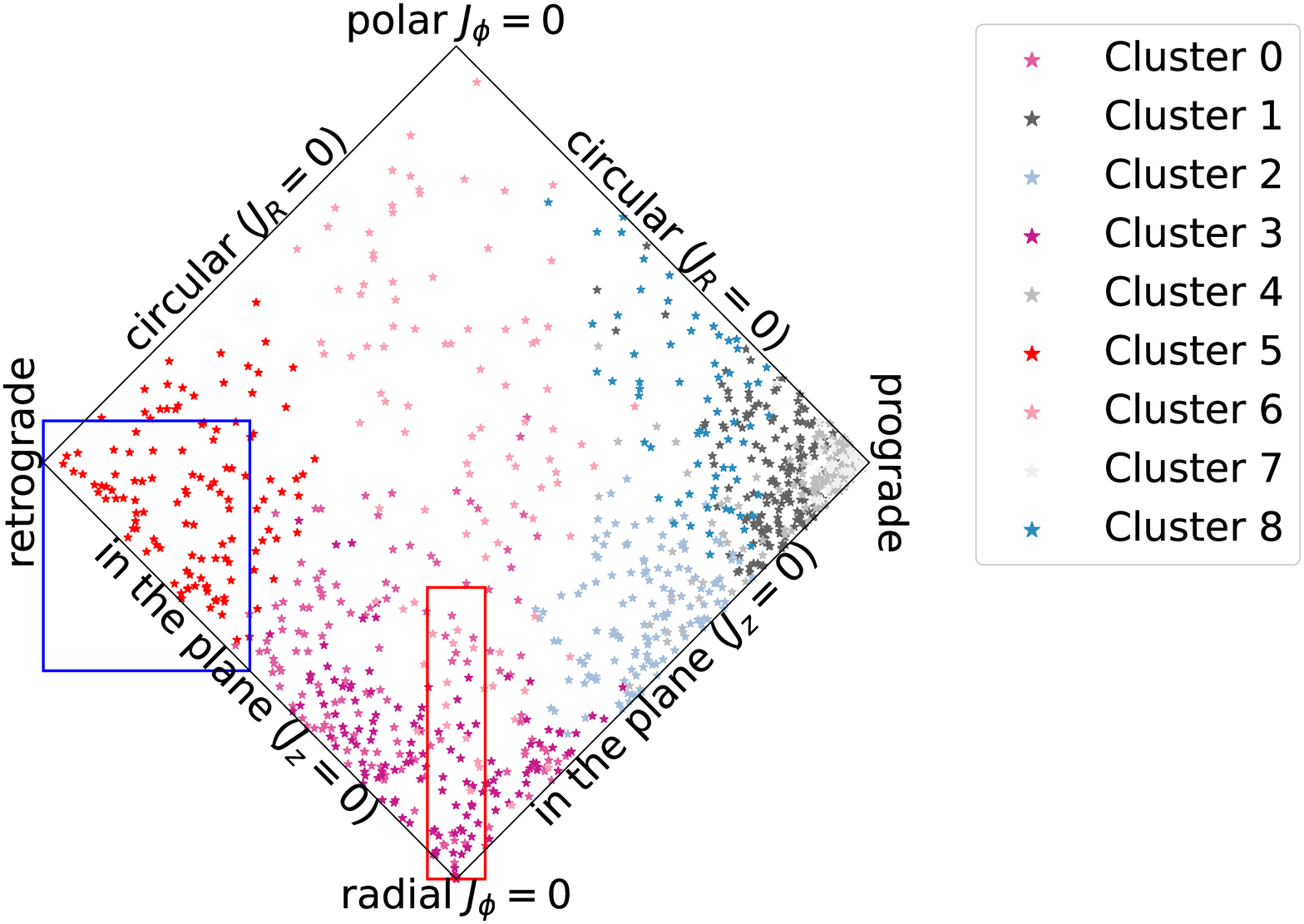}{fig2}{Action angles by total action. The blue box represents the limits of Sequoia in \citet{Myeong2018} and the red box the limits of GES in \citet{Myeong2018}. Cluster 5 (in red) agrees with  \citet{Myeong2018}.}

\articlefigure[width=.8\textwidth]{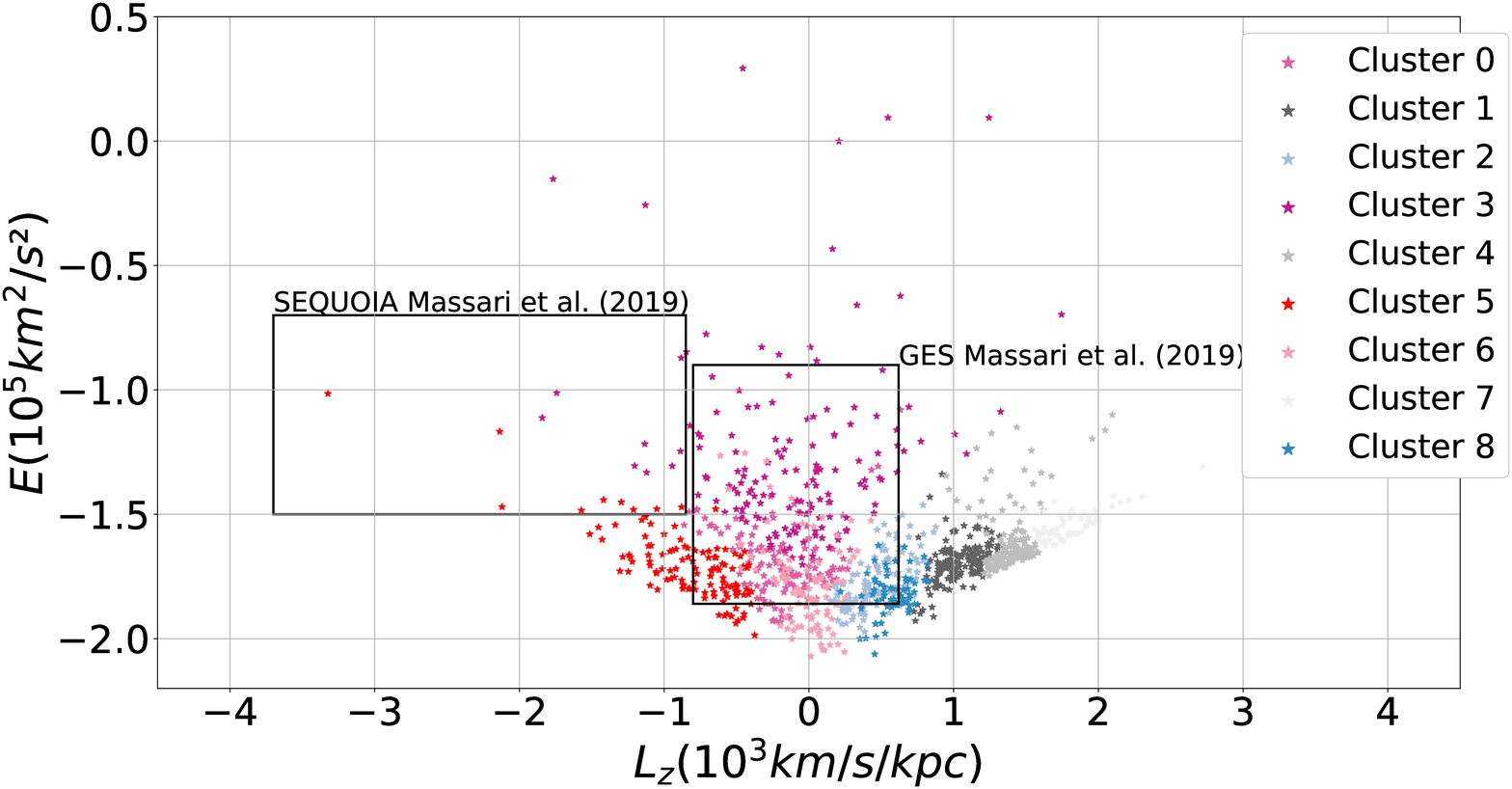}{fig3}{Total orbital energy by angular momentum in the Galactic pole direction (Lindblad diagram). Points colored in intermediate pink could be Wukong as described by \citet{Naidu2020}.}

\articlefigure[width=.8\textwidth]{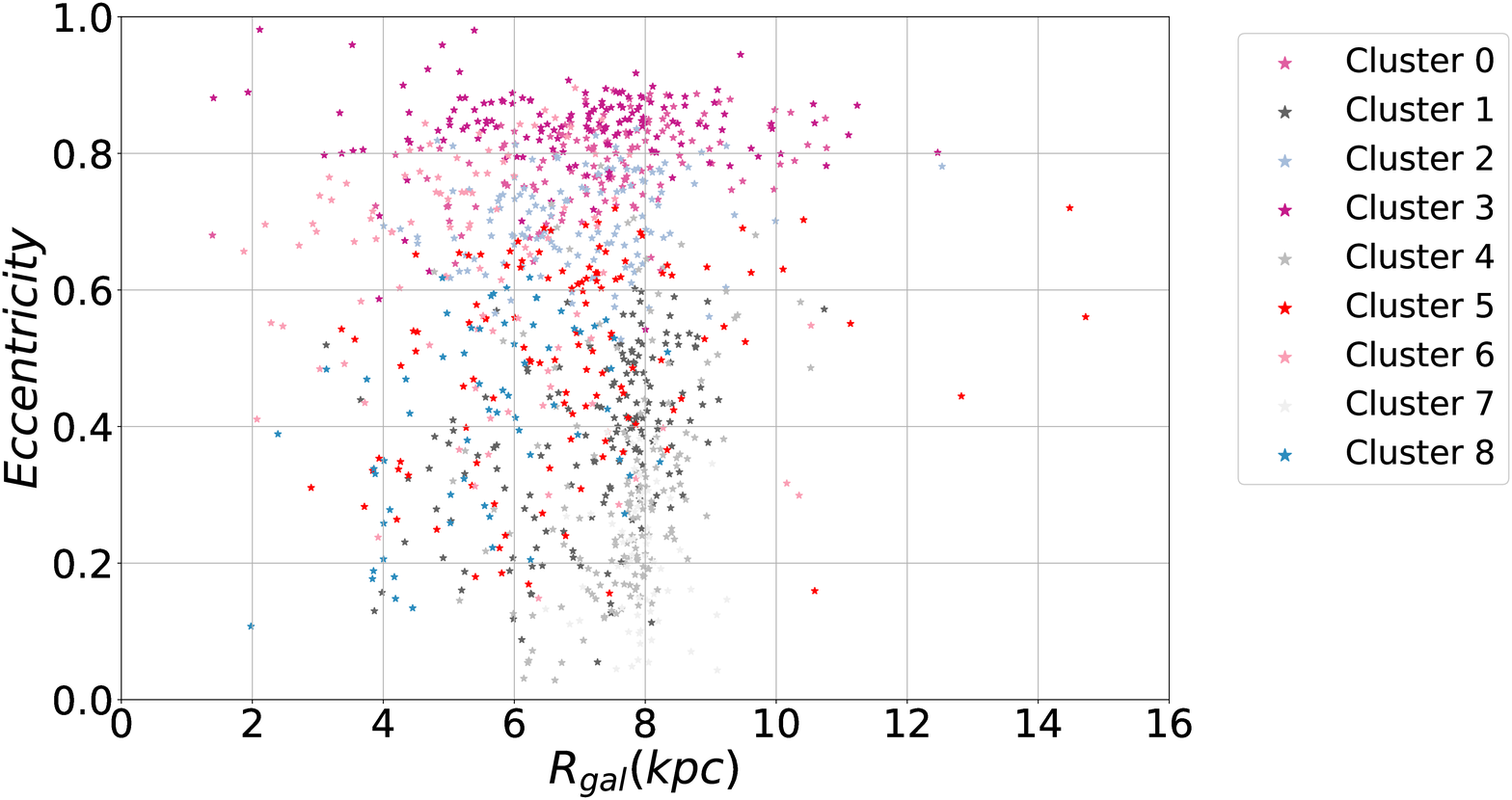}{fig4}{Orbital eccentricity by the average cylindrical radius of the orbit. The GES stars are mostly concentrated in high eccentricity.}
%\articlefiguretwo{P5-177_f1.eps}{P5-177_f2.eps}{fig1}{\emph{Left}:  \emph{Right}: }.

\acknowledgements A.R.S, R.S., and R.E.G. acknowledge support by the Polish National Science Centre through project 2018/31/B/ST9/01469.

\bibliographystyle{asp2014}
\bibliography{P5-177}
\end{document}